\documentclass{PoS}

	\usepackage{braket}
	\usepackage{cancel}
	\usepackage{graphicx}
	\usepackage{amssymb}
	\usepackage{amsmath}
	\usepackage{mathrsfs}
	\usepackage{xcolor}
	\usepackage{dcolumn}
	\usepackage{color}
	\usepackage{multirow}
	\usepackage{dcolumn}
	\usepackage[normalem]{ulem}
	\usepackage{bm}
	\usepackage{yhmath}
	\usepackage{booktabs}
	\usepackage{subfigure}
	\usepackage[utf8]{inputenc}

	\newcommand{\rcite}[1]{\cite{#1}}
	
	\newcommand{\refref}[1]{Ref.~\rcite{#1}}
	\newcommand{\eref}[1]{Eq.~(\ref{#1})}
	\newcommand{\sref}[1]{Sec.~(\ref{#1})}
	
	\newcommand{\fref}[1]{Fig.~(\ref{#1})}
	\newcommand{\nn}{\nonumber}
	
	\newcommand{\dkt}{}
	\newcommand{\dkf}{}
	\newcommand{\dxt}{}
	\newcommand{\dxf}{}

\title{Partial Wave Mixing in Hamiltonian Effective Field Theory}

\ShortTitle{Partial Wave Mixing in HEFT}

\author{
	\speaker{Yan Li}$^a$, Jia-jun Wu$^{a}$, Curtis D. Abell$^{b}$, Derek B. Leinweber$^{b}$ and Anthony W. Thomas$^{b,c}$\\ 
	\llap{$^a$}School of Physical Sciences, University of Chinese Academy of Sciences (UCAS), Beijing 100049, China\\
	\llap{$^b$}Special Research Center for the Subatomic Structure of Matter (CSSM), Department of Physics, University of Adelaide, Adelaide 5005, Australia\\
	\llap{$^c$}ARC Centre of Excellence for Particle Physics at the Terascale (CoEPP), Department of Physics, University of Adelaide, Adelaide, South Australia 5005, Australia
	\\ E-mail: \email{liyan175@mails.ucas.edu.cn}}

\abstract{
	Within general partial-wave mixing, a method for reducing the high dimension of the finite-volume Hamiltonian from Hamiltonian effective field theory is proposed. This method provides a new viewpoint on partial-wave mixing, and a set of matrices that can reflect the degree of partial-wave mixing. An example of isospin-2 $\pi\pi$ scattering is used to examine the consistency between this method and L\"{u}scher's method.}

\FullConference{37th International Symposium on Lattice Field Theory - Lattice2019\\
		16-22 June 2019\\
		Wuhan, China}

\begin{document}

\section{Introduction}
	The infinite-volume phase shift observed in experiment and the finite-volume spectrum obtained from Lattice QCD proved to be related by the model independent L\"{u}scher's formula \cite{Luscher:1985dn,Luscher:1986pf,Luscher:1990ux}.
	In most cases, a fitting process is necessary when applying L\"{u}scher's formula. For example, the phase shift is parameterized as the expansion of the momentum in \refref{Dudek:2012gj}.

	An equivalent approach is Hamiltonian effective field theory (HEFT) in which a potential is parameterized.
	HEFT was first introduced in \refref{Hall:2013qba} to study a $\Delta
	\to N\pi$ system, and was developed further in a series of
	works\cite{Hall:2014uca,Wu:2014vma,Liu:2015ktc,Liu:2016uzk,Liu:2016wxq,Wu:2016ixr,Wu:2017qve}.
	In the previous works, HEFT has only been applied to the cases without the inclusion of higher partial waves.
	If including the high partial wave contributions, the dimension of the finite volume Hamiltonian will increase fast. 
	Here a method for reducing the high dimension of the Hamiltonian within general partial-wave mixing is proposed. 
	It provides a new viewpoint on partial-wave mixing based on HEFT and the P-Matrix defined in \eref{eq:defPM} to reflect the degree of partial-wave mixing. 
	Finally, an example of isospin-2 $\pi\pi$ scattering is used to examine the consistency between this method and L\"{u}scher's method.

\section{Partial Wave Mixing in HEFT}
	\renewcommand{\dkt}[1]{\frac{d^3\mathbf{#1}}{(2\pi)^3}}
	\renewcommand{\dkf}[1]{\frac{d^4 #1}{(2\pi)^4}}
	\renewcommand{\dxt}[1]{d^3\mathbf{#1}~}
	\renewcommand{\dxf}[1]{d^4#1~}

	In the infinite volume, the spherical symmetry is described by the $O(3)$ group, while in the finite volume, the cubic symmetry is described by the $O_\text{h}$ group. The irreducible representations (irreps) of $O(3)$, the partial wave representation, are different from the irreps of $O_\text{h}$. They are related via the restricted representation as follows
	\begin{align}\label{eq:SFV-l2Gamma}
	  \mathbf{0}^+ &= \mathbf{A}_1^+ \,, \nn\\
	  \mathbf{1}^- &= \mathbf{T}_1^- \,, \nn\\
	  \mathbf{2}^+ &= \mathbf{E}^+ \oplus \mathbf{T}_2^+ \,, \nn\\
	  \mathbf{3}^- &= \mathbf{A}_2^- \oplus \mathbf{T}_1^- \oplus \mathbf{T}_2^- \,, \nn\\
	  \mathbf{4}^+ &= \mathbf{A}_1^+ \oplus \mathbf{E}^+ \oplus \mathbf{T}_1^+ \oplus \mathbf{T}_2^+ \,.
	\end{align}
	That different partial waves, decoupled in the infinite volume, can be mixed in the finite volume, is called partial-wave mixing.

	\subsection{Infinite- and Finite-Volume Hamiltonians}
		The infinite- and finite-volume Hamiltonians are given by
		\begin{align}
			\hat{H} &= \hat{H}_0 + \hat{V} = \int \dkt{k}\,h(k) \ket{\mathbf{k}}\bra{\mathbf{k}} + \int \dkt{p}\dkt{k} \,V(\mathbf{p},\mathbf{k}) \ket{\mathbf{p}}\bra{\mathbf{k}} \,, \nn\\
  			\hat{H}_L &= \hat{H}_{0L} + \hat{V}_L = \sum_{\mathbf{n}\in\mathbb{Z}^3} h\left(\frac{2\pi n}{L}\right) \ket{\mathbf{n}}\bra{\mathbf{n}} + \sum_{\mathbf{n}',\mathbf{n}\in\mathbb{Z}^3} V_L\left(\frac{2\pi\,\mathbf{n}'}{L},\,\frac{2\pi\,\mathbf{n}}{L}\right) \ket{\mathbf{n}'}\bra{\mathbf{n}} \,,
		\end{align}
		where $\ket{\mathbf{k}}$ and $\ket{\mathbf{n}}$ are the infinite-volume plane wave state and the finite-volume periodic plane wave state respectively, $h$ is the total kinematic energy, and $V(\mathbf{p},\mathbf{k})$ and $V_L(2\pi\,\mathbf{n}'/L,\,2\pi\,\mathbf{n}/L)$ are the infinite- and finite-volume potentials respectively. The normalization conditions are chosen to be
		\begin{equation}
			\braket{\mathbf{p}|\mathbf{k}} = (2\pi)^3\,\delta^3(\mathbf{p-k}) \qquad \braket{\mathbf{n}'|\mathbf{n}} 
              = \delta_{\mathbf{n}',\mathbf{n}} \,,
		\end{equation}
		in which case the infinite- and finite-volume potentials will be related by $V_L=V/L^3$.
		Normally, we have the following partial wave expansion for the potential
		\begin{equation}
			V(\mathbf{p},\mathbf{k}) = \sum_{l,m}\, v_l(p,k)\, Y_{lm}(\hat{\mathbf{p}})\, Y_{lm}^*(\hat{\mathbf{k}}) \,.
		\end{equation}
		
		The above discussion means that the infinite- and finite-volume Hamiltonians can be parameterized by a common set of parameters. Through fitting the finite-volume Hamiltonian eigenvalues to the lattice spectrum, we can then use the infinite-volume Hamiltonian to predict the scattering phase shift. This provides an equivalent approach to relate the spectrum and the phase shift as L\"{u}scher's formula.

	\subsection{How is the Dimension Reduced}

		Working with the original periodic plane wave basis $\ket{\mathbf{n}}$, we need to deal with $C_3(N)$ states at a lattice momentum sphere with $\mathbf{n}^2=N$. With the momentum cutoff $N_{\text{cut}}=600$, there are $\sum_{N=0}^{600}C_3(N) = 61,565$ states in total.
		
		To utilize the infinite-volume spherical symmetry, we combine $\ket{\mathbf{n}}$ with spherical harmonics to define
		\begin{equation}\label{eq:SFV-defnlm}
		  \ket{N;l,m} = \sum_{|\mathbf{n}|^2=N}\sqrt{4\pi}\, Y_{lm}(\hat{\mathbf{n}})\ket{\mathbf{n}} \,.
		\end{equation}
		Then with a partial wave cutoff $l_{\text{cut}}$ to ignore contributions from higher partial waves $l> l_{\text{cut}}$, we only need to deal with $(l_{\text{cut}}+1)^2$ states at any lattice momentum sphere. With $l_{\text{cut}}=4$, there are $600\times25+1$ states in total, where the $1$ comes from the state $\ket{\mathbf{n}=(0,0,0)}$.
		
		From the definition, the states $\ket{N;l,m}$ will inherit the rotational behavior from the spherical harmonics to behave as the vectors of the restricted representation, and hence will reduce into the states that respect the cubic symmetry as follows
		\begin{equation}
			\ket{N,l;\Gamma,f,\alpha} = \sum_m [C_l]_{\Gamma,f,\alpha;m}\ket{N;l,m} \,,
		\end{equation}
		where $(\Gamma,f,\alpha)$ represents the $\alpha$-th vector of the $f$-th occurrence (in a $l$) of the irrep $\Gamma$, and $[C_l]_{\Gamma,f,\alpha;m}$ are group theoretic constants independent of $N$.
		Considering the irrep $\Gamma=\mathbf{A}_1^+$, we should deal with only $2$ states at any lattice momentum sphere, as $\mathbf{A}_1^+$ occurs twice within $l_{\text{cut}}$ as shown in \eref{eq:SFV-l2Gamma}, so there are $600\times2+1$ states in total.

		Finally, we note the states $\ket{N,l;\Gamma,f,\alpha}$ are not orthonormalized, and can even be linear dependent. The orthonormalization will further reduce the dimension. In the isospin-2 $\mathbf{\pi\pi}$ scattering that will be discussed in \sref{sec:EIS}, for instance, $600\times2+1$ will be reduced to $923\,(\mathbf{A}_1^+)$, $965\,(\mathbf{E}^+)$ and $963\,(\mathbf{T}_2^+)$.

	\subsection{P-Matrix: Reflecting the Degree of Partial Wave Mixing}
		\begin{figure*}
			\centering
			\includegraphics[width=\textwidth]{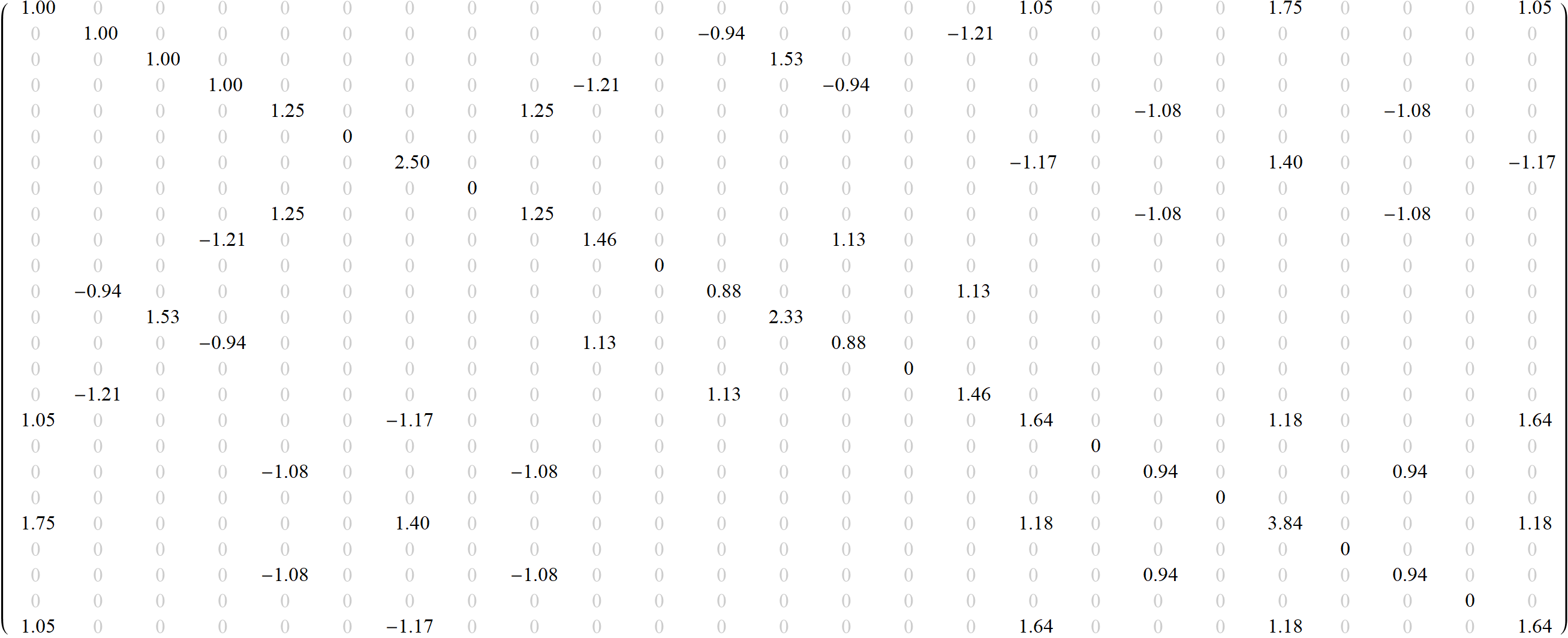}

			\vspace{.5cm}

			\includegraphics[width=\textwidth]{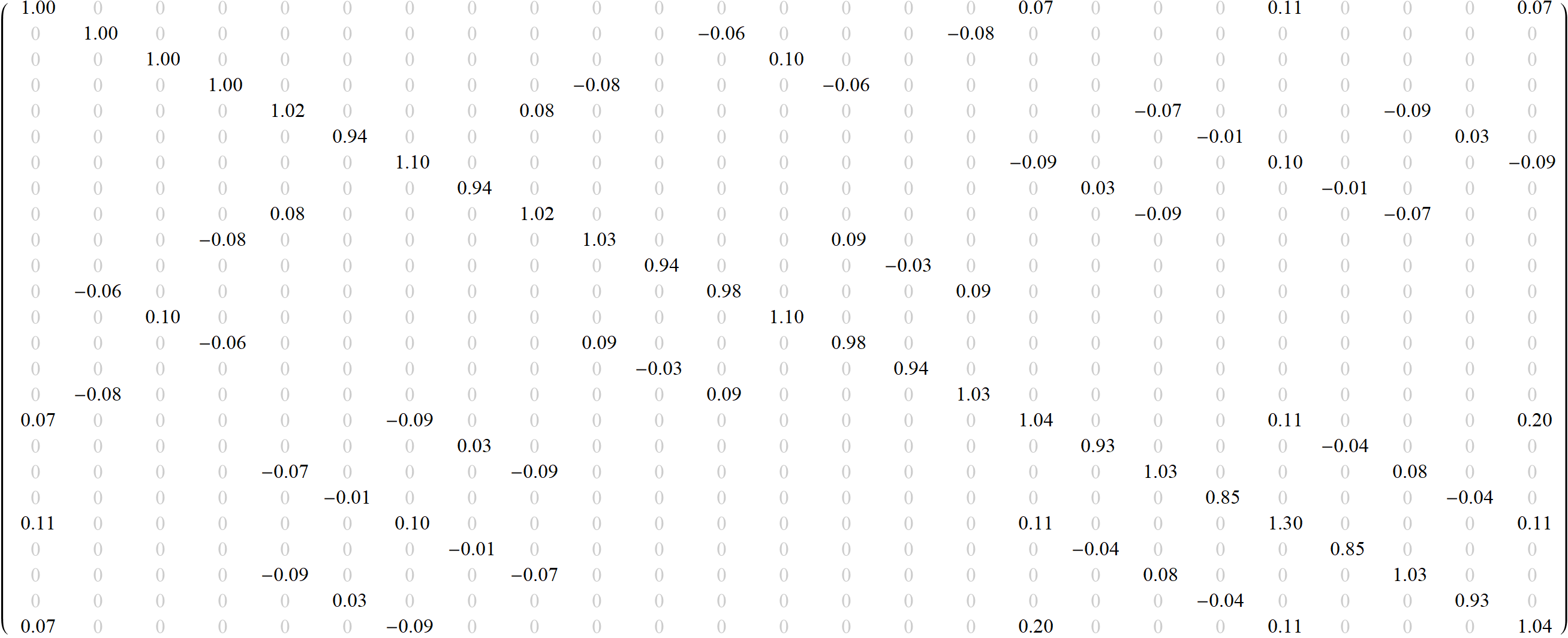}

			\vspace{.5cm}

			\includegraphics[width=\textwidth]{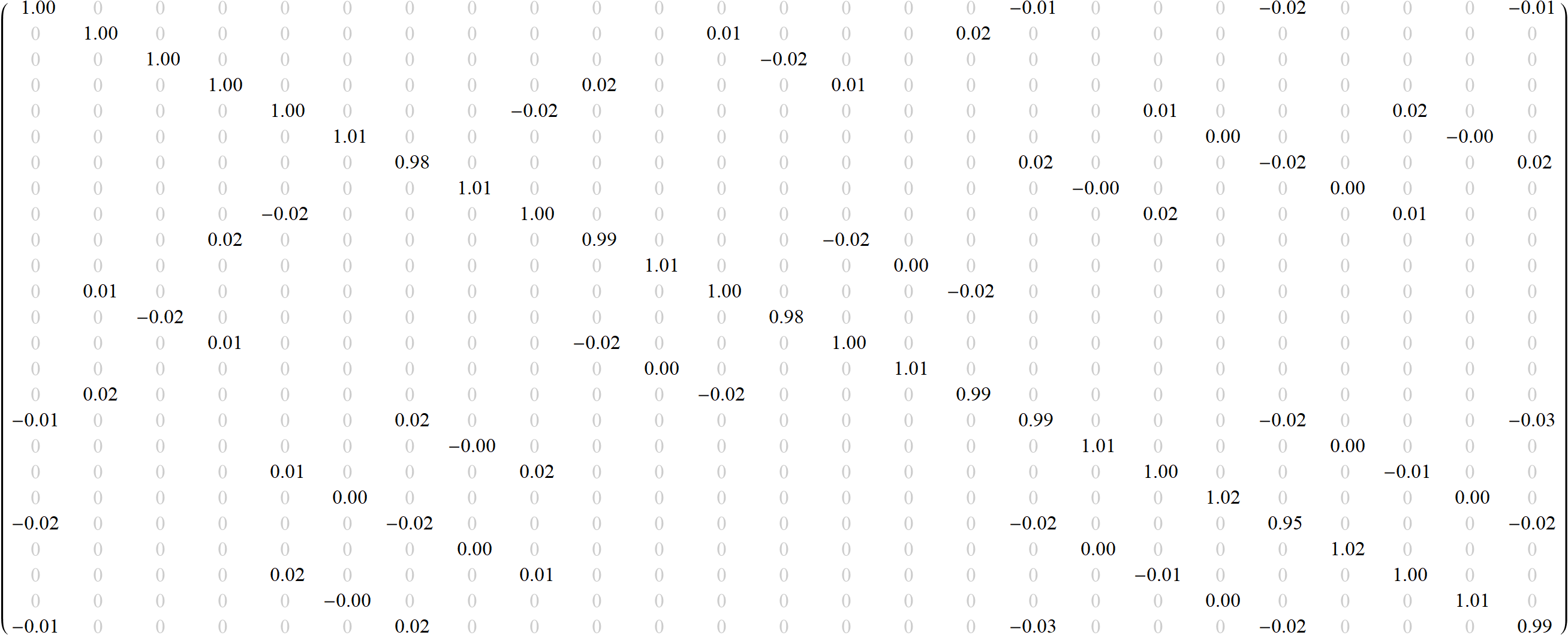}
			\caption{$P_{N}/C_3(N)$ with $N=1,\,581,\,941$ and $C_3(N)=6,\,336,\,552$ respectively: The $25\times25$ matrix ordered as $(l,m) =
			  (0,0),\,(1,-1),\,(1,0),\,(1,1),\,\cdots,\,(4,4).$}
			\label{fig:PM}
		\end{figure*} 

		To do the orthonormalization, the inner products between the states $\ket{N,l;\Gamma,f,\alpha}$ are needed. They will relate to the inner products between $\ket{N;l,m}$ via
		\begin{equation}\label{eq:PNLG}
			\braket{N,l';\Gamma,f',\alpha|N,l;\Gamma,f,\alpha} = \sum_{m',m}\, [C_{l'}]^*_{\Gamma,f',\alpha;m'}\,[P_{N}]_{l',m';l,m}\, [C_l]_{\Gamma,f,\alpha;m}\,,
		\end{equation}
		where we define the P-Matrix as the inner product matrix of the states $\ket{N;l,m}$ as follows
		\begin{equation}\label{eq:defPM}
			[P_N]_{l',m';l,m} := \braket{N;l',m'|N;l,m} = 4\pi \sum_{|\mathbf{n}|^2=N} Y_{l'm'}^*(\hat{\mathbf{n}})Y_{lm}(\hat{\mathbf{n}}) \,.
		\end{equation}
		The P-Matrix reflects the overlap between states with different partial wave quantum numbers $(l,m)$, and hence can reflect the degree of partial-wave mixing.

		From the P-Matrix, we can see the recovery of spherical symmetry in the infinite-volume limit.
		When $L\to\infty$, if we care about a finite and non-zero momentum $2\pi\sqrt{N}/L$, we will have $N\to\infty$. The summation at lattice momentum sphere $\sum_{|\mathbf{n}|^2=N}$ will then approximate an integration over solid angle, and the property of spherical harmonics will tell us
		\begin{equation}
			[P_N]_{l',m';l,m} \to C_3(N)\,\delta_{l',l}\,\delta_{m',m} \,,
		\end{equation}
		where the factor $C_3(N)$ comes from the consideration of $[P_N]_{0,0;0,0}=C_3(N)$.
		Here we provide a few examples of the numerical values for $P_N/C_3(N)$ with $N = 1$, $581$ and $941$ in \fref{fig:PM}, and it is indeed approaching an identity matrix.

		The P-Matrix also respects the cubic symmetry. As indicated in \eref{eq:PNLG}, the P-Matrix can be made block diagonal according to the irreps of the cubic group by a unitary transformation implemented via a $N$-independent matrix constructed from the constants $[C_l]_{\Gamma,f,\alpha;m}$.

\section{Example of Isospin-2 {\large $\mathbf{\pi\pi}$} Scattering}\label{sec:EIS}
\begin{figure*}[h]
		\centering
		\includegraphics[width=7cm]{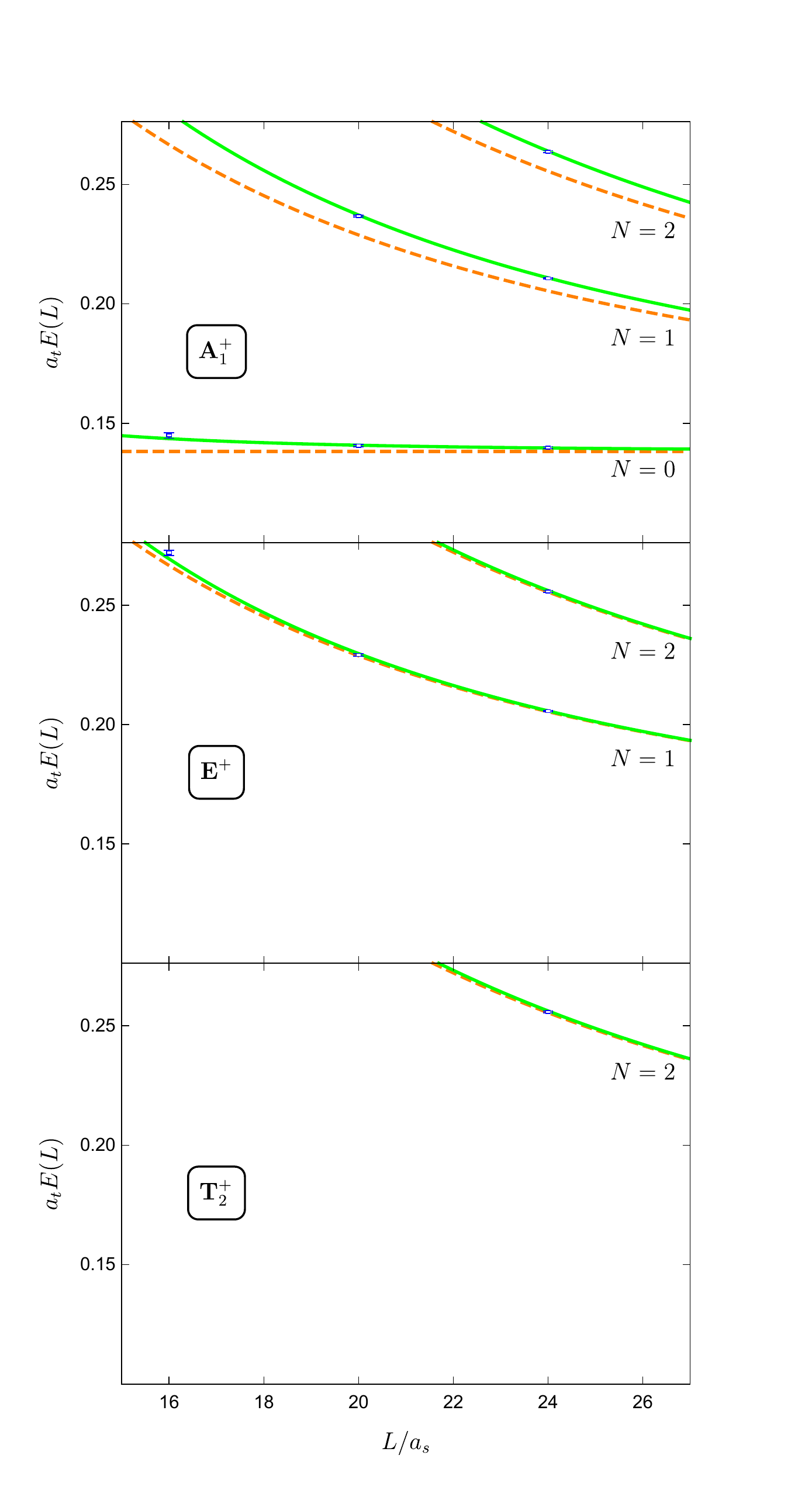}
		\includegraphics[width=7cm]{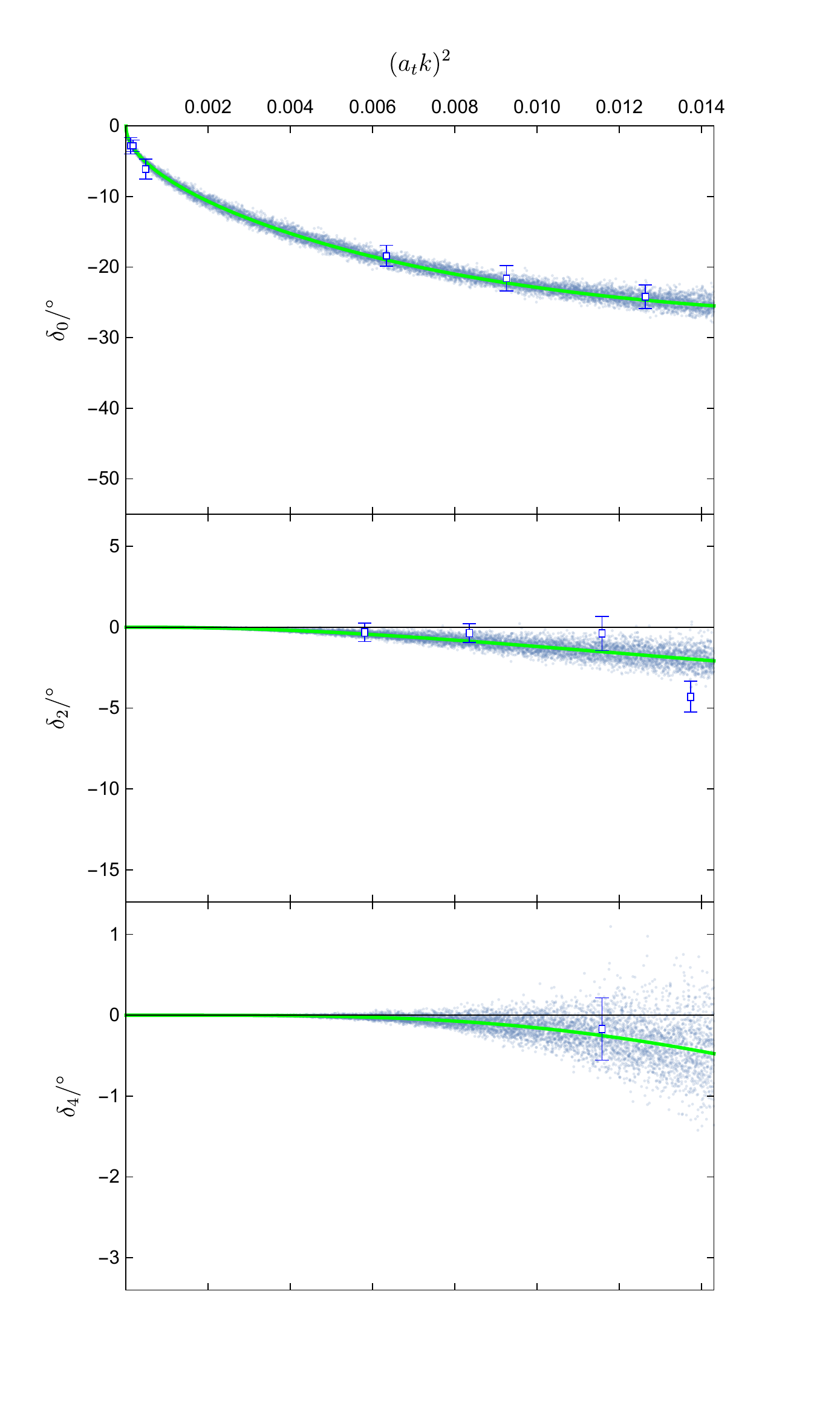}
		\caption{Left: Data points describe the lattice spectrum for irreps $\mathbf{A}_1^+$, $\mathbf{E}^+$, $\mathbf{T}_2^+$ -- from \refref{Dudek:2012gj}. Dashed curves represent the non-interacting rest-frame pion-pair energies $2\sqrt{m_\pi^2+k_N^2}$. Solid curves represent the HEFT prediction of the volume dependent spectrum using the fitted parameters. Right: Data points are phase shifts predicted by L\"{u}scher's method -- from \refref{Dudek:2012gj}. Solid curves represent the HEFT prediction of the $s$- (top), $d$- (middle) and $g$- (bottom) wave phase shifts and the scattered points describe the uncertainty.}
		\label{fig:Ex}
	\end{figure*}

	Here we choose the following parametrization for the potential (measured in units of lattice spacings)
	\begin{equation}
		v_l(p,k)=f_l(p)\,G_l\,f_l(k) \,,\qquad f_l(k) = \frac{(d_l\,k)^{l}}{\left(1+(d_l\,k)^2\right)^{l/2+2}} \,,
	\end{equation}
	where there are 2 parameters $G_l$ and $d_l$ for each partial wave.
	With a partial wave cutoff $l_{\text{cut}}=4$, Bose symmetry only allows $s$-, $d$- and $g$-waves. The lattice QCD results are from \refref{Dudek:2012gj}. Our HEFT results are shown in \fref{fig:Ex}, in which HEFT and L\"{u}scher's method are indeed consistent.		

\section{Summary and Outlook}
	We have discussed the finite-volume partial-wave mixing in HEFT. We showed how the dimension of the Hamiltonian is reduced. We defined the P-Matrix which can reflect the degree of partial-wave mixing. Then an example of isospin-2 $\pi\pi$ scattering was used to check the consistency between HEFT and L\"{u}scher's method.

	In future work, the moving system formalism of HEFT will be developed. The pion mass dependence of a system with a resonance will be explored in HEFT.

%----------------------------------------------------------------------------------------
\section*{ACKNOWLEDGEMENTS}
	It is a pleasure to thank Stephen Sharpe for interesting discussions on the research presented
	herein during his visit as a George Southgate Fellow.  The finite-volume energy levels and their
	covariances from \refref{Dudek:2012gj} were provided by the Hadron Spectrum Collaboration -- no
	endorsement on their part of the analysis presented in the current paper should be assumed.
	This project is also supported by the Thousand Talents Plan for Young Professionals.
	This
	research was supported by the Australian Research Council through ARC Discovery Project Grants
	Nos.\ DP150103101 and DP180100497 (A.W.T.) and DP150103164 and DP190102215 (D.B.L.).

\bibliographystyle{JHEP}
\bibliography{references}

\providecommand{\href}[2]{#2}\begingroup\raggedright\begin{thebibliography}{10}

\bibitem{Luscher:1985dn}
M.~Lüscher, \emph{{Volume Dependence of the Energy Spectrum in Massive Quantum
  Field Theories. 1. Stable Particle States}},
  \href{https://doi.org/10.1007/BF01211589}{\emph{Commun. Math. Phys.}
  {\bfseries 104} (1986) 177}.

\bibitem{Luscher:1986pf}
M.~Lüscher, \emph{{Volume Dependence of the Energy Spectrum in Massive Quantum
  Field Theories. 2. Scattering States}},
  \href{https://doi.org/10.1007/BF01211097}{\emph{Commun. Math. Phys.}
  {\bfseries 105} (1986) 153}.

\bibitem{Luscher:1990ux}
M.~Lüscher, \emph{{Two particle states on a torus and their relation to the
  scattering matrix}},
  \href{https://doi.org/10.1016/0550-3213(91)90366-6}{\emph{Nucl. Phys.}
  {\bfseries B354} (1991) 531}.

\bibitem{Dudek:2012gj}
J.~J. Dudek, R.~G. Edwards and C.~E. Thomas, \emph{{S and D-wave phase shifts
  in isospin-2 pi pi scattering from lattice QCD}},
  \href{https://doi.org/10.1103/PhysRevD.86.034031}{\emph{Phys. Rev.}
  {\bfseries D86} (2012) 034031}
  [\href{https://arxiv.org/abs/1203.6041}{{\ttfamily 1203.6041}}].

\bibitem{Hall:2013qba}
J.~M.~M. Hall, A.~C.~P. Hsu, D.~B. Leinweber, A.~W. Thomas and R.~D. Young,
  \emph{{Finite-volume matrix Hamiltonian model for a $\Delta \to N\pi$
  system}}, \href{https://doi.org/10.1103/PhysRevD.87.094510}{\emph{Phys. Rev.}
  {\bfseries D87} (2013) 094510}
  [\href{https://arxiv.org/abs/1303.4157}{{\ttfamily 1303.4157}}].

\bibitem{Hall:2014uca}
J.~M.~M. Hall, W.~Kamleh, D.~B. Leinweber, B.~J. Menadue, B.~J. Owen, A.~W.
  Thomas et~al., \emph{{Lattice QCD Evidence that the $\Lambda$(1405) Resonance
  is an Antikaon-Nucleon Molecule}},
  \href{https://doi.org/10.1103/PhysRevLett.114.132002}{\emph{Phys. Rev. Lett.}
  {\bfseries 114} (2015) 132002}
  [\href{https://arxiv.org/abs/1411.3402}{{\ttfamily 1411.3402}}].

\bibitem{Wu:2014vma}
J.-J. Wu, T.~S.~H. Lee, A.~W. Thomas and R.~D. Young, \emph{{Finite-volume
  Hamiltonian method for coupled-channels interactions in lattice QCD}},
  \href{https://doi.org/10.1103/PhysRevC.90.055206}{\emph{Phys. Rev.}
  {\bfseries C90} (2014) 055206}
  [\href{https://arxiv.org/abs/1402.4868}{{\ttfamily 1402.4868}}].

\bibitem{Liu:2015ktc}
Z.-W. Liu, W.~Kamleh, D.~B. Leinweber, F.~M. Stokes, A.~W. Thomas and J.-J. Wu,
  \emph{{Hamiltonian effective field theory study of the $\mathbf{N^*(1535)}$
  resonance in lattice QCD}},
  \href{https://doi.org/10.1103/PhysRevLett.116.082004}{\emph{Phys. Rev. Lett.}
  {\bfseries 116} (2016) 082004}
  [\href{https://arxiv.org/abs/1512.00140}{{\ttfamily 1512.00140}}].

\bibitem{Liu:2016uzk}
Z.-W. Liu, W.~Kamleh, D.~B. Leinweber, F.~M. Stokes, A.~W. Thomas and J.-J. Wu,
  \emph{{Hamiltonian effective field theory study of the $\mathbf{N^*(1440)}$
  resonance in lattice QCD}},
  \href{https://doi.org/10.1103/PhysRevD.95.034034}{\emph{Phys. Rev.}
  {\bfseries D95} (2017) 034034}
  [\href{https://arxiv.org/abs/1607.04536}{{\ttfamily 1607.04536}}].

\bibitem{Liu:2016wxq}
Z.-W. Liu, J.~M.~M. Hall, D.~B. Leinweber, A.~W. Thomas and J.-J. Wu,
  \emph{{Structure of the $\mathbf{\Lambda(1405)}$ from Hamiltonian effective
  field theory}}, \href{https://doi.org/10.1103/PhysRevD.95.014506}{\emph{Phys.
  Rev.} {\bfseries D95} (2017) 014506}
  [\href{https://arxiv.org/abs/1607.05856}{{\ttfamily 1607.05856}}].

\bibitem{Wu:2016ixr}
J.-J. Wu, H.~Kamano, T.~S.~H. Lee, D.~B. Leinweber and A.~W. Thomas,
  \emph{{Nucleon resonance structure in the finite volume of lattice QCD}},
  \href{https://doi.org/10.1103/PhysRevD.95.114507}{\emph{Phys. Rev.}
  {\bfseries D95} (2017) 114507}
  [\href{https://arxiv.org/abs/1611.05970}{{\ttfamily 1611.05970}}].

\bibitem{Wu:2017qve}
J.-j. Wu, D.~B. Leinweber, Z.-w. Liu and A.~W. Thomas, \emph{{Structure of the
  Roper Resonance from Lattice QCD Constraints}},
  \href{https://doi.org/10.1103/PhysRevD.97.094509}{\emph{Phys. Rev.}
  {\bfseries D97} (2018) 094509}
  [\href{https://arxiv.org/abs/1703.10715}{{\ttfamily 1703.10715}}].

\end{thebibliography}\endgroup

\end{document}